\documentclass[12pt]{article}
\usepackage{amsmath}
\usepackage{graphicx}
\usepackage[latin1]{inputenc}

\begin{document}
	\author{Ram Brustein, Yoav Zigdon}
	\title{\vspace{-1.8in}
		{Revealing the interior of black holes out of equilibrium in the SYK model}}
	\author{{\large Ram Brustein, Yoav Zigdon}
		\\
		\vspace{-.5in} \hspace{-0.05in} \vbox{\normalsize
			\begin{center}
				\ Department of Physics, Ben-Gurion University, \\
				Beer-Sheva 84105, Israel
				\	\\ \small \hspace{0.07in}
				ramyb@bgu.ac.il,\ yoavzig@post.bgu.ac.il
			\end{center}
		}}
		\date{}
		\maketitle
		
		\begin{abstract}
			The Sachdev-Ye-Kitaev (SYK) model can be used to describe black holes (BHs) in two-dimensional nearly-Anti-de-Sitter gravity. We show that when such BHs are perturbed by a time-dependent negative-energy perturbation, their interior can be partially revealed.  In the SYK model, a partial measurement of the state of the Majorana fermion pairs allows one to construct a matching time-dependent negative-energy perturbation of the BH geometry that shifts the state of the BH away from its equilibrium state. Kourkoulou and Maldacena  showed that, if the perturbation is strong enough, the interior can be fully exposed and the BH disappears. Here, we show that when the perturbation is weaker than the threshold for full exposure, it effectively moves the horizon of the BH inwards, thus partially exposing the interior of the BH and leaving behind a smaller BH. The exposure is in proportion to the number of measured Majorana pairs and so also in proportion to the magnitude of the energy of the perturbation. From the boundary, the partial measurement is perceived as a burst of radiation whose strength and duration are proportional to the number of measured Majorana pairs.
		\end{abstract}
		
		\newpage
		\section{Introduction}
		In classical general relativity (GR), the black hole (BH) interior is an empty region with (possibly) some singular core. Quantum theory suggests otherwise \cite{AMPS,AMPS2,Mathur1,Sunny,Braunstein}. Yet, it is unclear whether and how it is possible for an observer outside the BH to determine the state of the interior.  The BH is surrounded by a  horizon which, classically, prevents such an observer from obtaining any information about the interior. Quantum mechanically, BHs in equilibrium emit Hawking radiation \cite{Hawking}. Because the rate of emission is so slow, an external observer has to wait for a Page time \cite{Page} until the state of the BH can start to be ``read". If all of the Hawking radiation is collected (a practically impossible task), the observer would be able to determine the state of the BH. From this perspective, one can view the evaporation process as gradually exposing the state of the BH interior.
		
		It was argued in \cite{QuantumHair} (see also \cite{BlackHolesCollide,discovering}) that, when the BH is out of equilibrium, it can emit ``supersized" Hawking radiation -- radiation of much larger amplitude than the standard Hawking radiation. Then some properties of the state of the BH can be determined much earlier than the Page time by monitoring the large-amplitude coherent emission of gravitational waves. This emitted radiation is analogous to stimulated emission, in contrast to the standard Hawking radiation which is analogous to spontaneous emission.
		
		What does it mean to observe the state of the interior? An outside observer must interpret any radiation coming out from the BH as emitted from a region outside the horizon. Hence, the stimulated Hawking emission must be viewed from an outside observer perspective as induced by a negative-energy perturbation that caused the horizon to recede inwards, thus exposing part of the BH interior to outside observations.
		
		In this paper we consider two-dimensional (2d) nearly-Anti-de Sitter  (NAdS${}_2$) BHs and, using the Sachdev-Ye-Kitaev (SYK) model, demonstrate the validity of the ideas which were just outlined by an explicit calculation.

		The SYK model \cite{SachdevYe,SYK} describes $N$ Majorana fermions interacting all-to-all with random coupling constants  whose typical strength is $J$. In  the large $N$ and the zero-temperature limit the model has a reparametrization symmetry. When the temperature is finite but small, the dynamics of this reparametrization mode is determined by the Schwarzian action \cite{SYK,Kitaev,Comments} (see also \cite{AdS2SYKreview}).
		
		This model is interesting because it is related to nearly-AdS${}_2$ gravity \cite{AdS2,Jensen,Engelsoy,MaldacenaAdS2,JT}, for which the fundamental symmetry is that of time reparametrizations in the limit of zero dilaton field (pure GR).
		
		The Schwarzian action describes the time reparametrization on the boundary for finite but small dilaton. The solution of the equation of motion (EOM) derived from the Schwarzian action describes a boundary which has a finite time extent. This bounded spacetime can be described by the exterior of an AdS${}_2$ Rindler BH, as will be shown in detail later on.

The significance of our work is not restricted to 2d gravity because four-dimensional (near-extremal) black holes can be described by the dilaton-gravity model that we have used. This is explained, for example, in \cite{AdS2SYKreview}: The entropy of such black holes has a leading correction proportional to the temperature, which is captured by the Schwarzian action. The near-extremality is associated with the AdS${}_2$ near-horizon geometry. Nevertheless, we do expect that our results are relevant to BH's out of equilibrium in general.
		
		Recently, Kourkoulou and Maldacena  (KM) \cite{MaldacenaSpins} considered a measurement of all Majorana pairs, which, from a spacetime point of view,  corresponds to a time-dependent, negative-energy perturbation which reveals the entire region behind the horizon of the NAdS${}_2$ BH. Thus KM showed that a strong enough perturbation corresponds to a measurement of the state of the BH interior. From a spacetime point of view, the BH disappears completely as a result of the perturbation and an outside observer can therefore fully determine the state of the interior.
		
		Here, we assume a measurement on the subsystem of a large number of Majorana pairs and show that the BH interior is partially revealed due to a perturbation which takes the BH out of equilibrium. The measurement is of an essential importance. If the pairs are not measured, the average value of the perturbation Hamiltonian vanishes. If the pairs are measured, this corresponds to a negative-energy perturbation that induces an inward shift of the horizon: some part of the geometry that was previously inside the horizon is outside the horizon after the perturbation is applied. The extent of the exposure is shown to be proportional to the number of measured Majorana pairs and also to the magnitude of the energy of the perturbation. We find that from the boundary point of view, the partial measurement is interpreted as a flow of energy in the form of radiation. A boundary observer can ascribe the source of this radiation to the interior region of the BH that exposed by the perturbation.

		The collision of two astrophysical BHs creates, as an intermediate state, a BH out of equilibrium which decays to its equilibrium state by emitting (mostly) gravitational waves.  Then, following the same logic as above, the interior of the created BH should be partially exposed to outside observations via its stimulated emission. Recently, several scenarios were discussed \cite{sc1,sc2,sc3,sc4,BlackHolesCollide,discovering} in which the emission from such BH's should be different than the one predicted by GR.  Therefore, our ideas could perhaps be tested by measuring the gravitational waves that they emit.
		
		The paper is organized as follows: First, we briefly review the SYK model in the low temperature regime $1 \ll\beta J \ll N$, the NAdS${}_2$ boundary and the relation to BHs in equilibrium.
		Then, we discuss BHs out of equilibrium.  We show that the ``mass-term'' perturbation corresponding to the measurement of a large number of Majorana pairs extends the boundary trajectory of NAdS${}_2$. We calculate the amount of extension in global coordinates. This is followed by a calculation of the corresponding inwards shift of the horizon and then the flux of energy that flows to the boundary.  The paper ends with a summary and interpretation of our results.

		\section{Description of black holes in equilibrium with the SYK model}
		
		The Sachdev-Ye-Kitaev (SYK) model \cite{SachdevYe,SYK} describes $N$ Majorana fermions interacting via a four-body interaction
		\begin{equation}
		H = \sum_{i<k<l<m} ^N j_{iklm} \psi_i \psi_k \psi_l \psi_m.
		\label{sykham}
		\end{equation}
		Here $\psi_i$ is the i'th Majorana field and $j_{iklm}$ are the  coupling constants, which are randomly distributed with zero mean and a variance of $\frac{3! J^2}{N^3}$. The constant $J$ sets the interaction strength.
		
		In the Infra-Red limit $\beta J =\infty$ ($\beta$ is the inverse temperature) and large number of Majorana fermions  $1\ll N$ the model has a reparametrization symmetry \cite{SYK,Kitaev,Comments}. When $\tau \to f(\tau)$, the two-point function of the model
		$
		G(\tau,\tau')=\frac{1}{N}\sum_{i=1} ^N \langle T\psi_i(\tau)\psi_i(\tau')\rangle
		$
		transforms as
		\begin{equation}
		\label{RepSymmetry}
		G(\tau,\tau') \to \left(f'(\tau)f'(\tau')\right)^{\frac{1}{4}}G(f(\tau),f(\tau')).
		\end{equation}
		
		We will consider the case of a large number of Majorana fermions and low-energies: $1\ll \beta J \ll N$. In this regime, the SYK model describes effectively a single degree of freedom $f(t)$ - the reparametrization, governed by the Schwarzian action \cite{SYK,Kitaev,Comments}:
		\begin{equation}
		\label{SchAction}
		S = -\frac{N \alpha_S}{J} \int dt \{f(t),t\},
		\end{equation}
		where $\alpha_S$ is a numerical constant and
		\begin{equation}
		\label{SchDerivative}
		\{ f(t),t \} =\left(\frac{f''}{f'}\right)'-\frac{1}{2}\left(\frac{f''}{f'}\right)^2.
		\end{equation}
		We now briefly review the AdS${}_2$ spacetime.  Three coordinate systems that shall be used with their respective metrics are presented in Table~1. The AdS length scale is denoted by $R$.
		\begin{table}[t]
			\hspace{-.45in}
			\begin{tabular}{|c|c|c|c|}
				\hline
				Coordinates& Temporal range & Spatial range & Geometry  \\ \hline
				Global &$ -\infty<\tau<\infty$ & $-\frac{\pi}{2} \leq \rho \leq \frac{\pi}{2}$ &
				$ \frac{ds^2}{R^2} = \frac{-d\tau^2 + d\rho^2}{\cos^2 \rho}$ \\ \hline
				Poincar\'e-Patch & $-\infty <T<\infty$ & $0<z<\infty$ & $\frac{ds^2}{R^2}= \frac{-dT^2+dz^2}{z^2}$ \\ \hline
				Rindler & $-\infty<t_R<\infty$ & $\sqrt{M}<\frac{r}{R}<\infty~$ &
				$ds^2 = -\frac{(r^2-R^2 M)}{R^2}dt_R^2+\frac{R^2 }{r^2-R^2 M}dr^2$  \\
				\hline
			\end{tabular}\\
			\caption{Coordinate systems in AdS${}_2$ spacetime.}
		\end{table}	
The position of the horizon in Rindler coordinates is set by a dimensionless parameter $M$: $r_h=R\sqrt{M}$. The relations between the different coordinate systems are:
		\begin{equation}
		\label{global}
		z=\frac{R \cos(\rho)}{\cos(\tau)+\sin(\rho)} ,~ T=\frac{R \sin(\tau)}{\cos(\tau)+\sin(\rho)}.
		\end{equation}
		\begin{equation}
		\label{rindler}
		\frac{r}{R\sqrt{M}}=\frac{\cos(\tau)}{\cos(\rho)} ~,~ \tanh\left(\frac{t_R\sqrt{M}}{R}\right)=\frac{\sin(\tau)}{\sin(\rho)}.
		\end{equation}
		Below, we briefly review the essentials of NAdS${}_2$ gravity  \cite{AdS2,Jensen,Engelsoy,MaldacenaAdS2} and its relation to the SYK model and BHs.

		The boundary of NAdS${}_2$ deviates from the boundary of AdS${}_2$, which is at $z=0$. It was shown in \cite{MaldacenaAdS2} that the boundary of NAdS${}_2$ can be parametrized as $(T(t),z(t)=\epsilon T'(t))$ where $T(t)$ is the Poincar\'e time coordinate  and $\epsilon$ is small.  The Schwarzian action describes the boundary value $T(t)$ of NAdS${}_2$ gravity,
		\begin{equation}
		\label{SchAction2}
		S = -\frac{1}{8\pi G} \bar{\phi}_r \int dt  \{T(t),t\},
		\end{equation}
		where $\bar{\phi}_r$ is a constant and $G$ is the Newton constant in 2D.
		
		Comparing action (\ref{SchAction}) and (\ref{SchAction2}), it follows that $f(t)$ should be identified with $T(t)$ and $f'(t)$ with $z(t)$.
		
		A solution of the EOM derived from Eq.~(\ref{SchAction}) is the following,
		\begin{equation}
		\label{sol1}
		f(t) = \frac{\pi}{J^2 \beta} \tanh\left(\frac{\pi t}{\beta} \right).
		\end{equation}
		The function $f(t)$ in Eq.~(\ref{sol1}) is bounded by ${\pi}/{(J^2 \beta)}$. Since $f(t)$ is identified with $T(t)$, it follows that the extent of time on the boundary of the NAdS${}_2$ is also bounded by an upper bound $T_{\text{bound}}$, i.e. $T(t) \leq T_{\text{bound}}$.
		
		\begin{figure}[t]
			\begin{center}
				\includegraphics[scale=0.25]{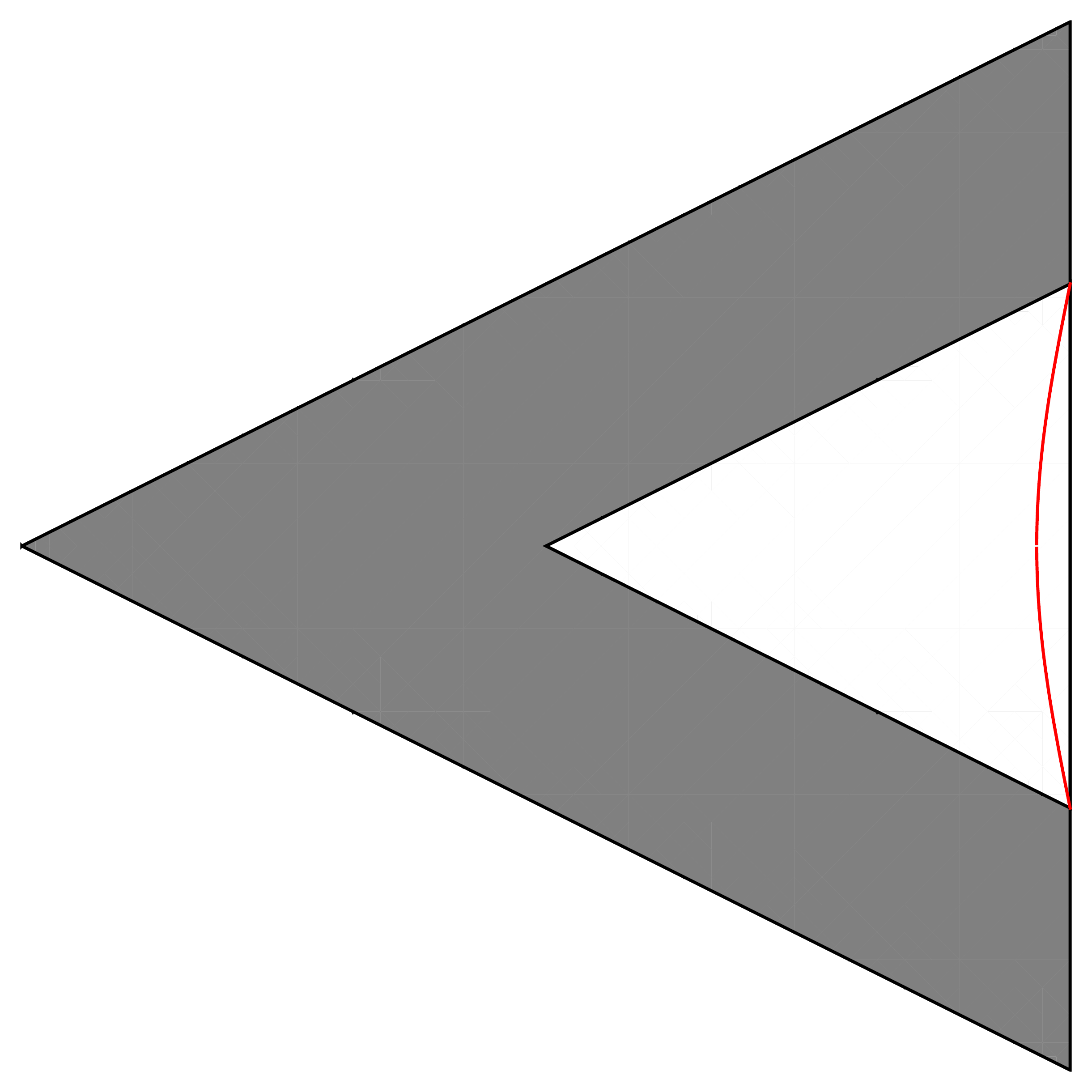}
			\end{center}
			\caption{The white region is accessible to the boundary observer whose world line is depicted by the (red) curve near the boundary. The shaded region describes the interior of the black hole, which is inaccessible to the observer. }
		\end{figure}

		One can compactify the spacetime:
		\begin{equation}
		\widetilde{T} = \frac{R}{T_{\text{bound}}} T ~;~ \widetilde{z} = \frac{R}{T_{\text{bound}}}z,
		\end{equation}
		therefore $\widetilde{T}\leq R$. In global coordinates (\ref{global}), $\tau\leq \frac{\pi}{2}$. Consider an observer on the NAdS${}_2$ boundary whose trajectory starts at  $\tau=-\frac{\pi}{2}$. The accessible region for this observer in the Penrose diagram is a triangle (see Figure 1). This region is spanned by the Rindler BH coordinates  ($t_R$, $r$), related to global coordinates as in Eq.~(\ref{rindler}). The horizon is given by the two lines $\tau=\pm \rho$.
		
		It follows that the NAdS${}_2$ spacetime is a BH spacetime, whose interior corresponds to the inaccessible region for the boundary observer. This is a two-dimensional Rindler BH in the Poincar\'e-Patch.

\section{Black holes out of equilibrium}
		
The purpose of this section is to introduce a time-dependent negative-energy perturbation to the SYK model. This perturbation depends on knowing the state of some (or all) of the Majorana pairs. We will show that when the BH is perturbed, a region that was previously behind the horizon becomes exposed to an outside observer.
		
		Maldacena and Kourkoulou \cite{MaldacenaSpins} considered the pure ``spin-states'' $|B_s\rangle$, defined by:
		\begin{equation}
		2i \psi_{2k-1} \psi_{2k} |B_s \rangle = s_k |B_s \rangle ~,~ s_k = \pm 1 ~,~ k=1,...,N/2
		\end{equation}
		and then applied Euclidean evolution with the Hamiltonian (\ref{sykham}) to them:
		\begin{equation}
		|B_s (\beta) \rangle = e^{-\beta H /2} |B_s \rangle.
		\end{equation}
		
		For $\tau=\frac{\beta}{2}+it$ and $1\ll \beta J \ll N$, they showed that
		\begin{equation}
		\label{Ofdiagonal}
		\langle B_s (\beta)| s_k \psi_{2k-1}(\tau) \psi_{2k}(\tau) |B_s (\beta)\rangle=-\frac{i\sqrt{\pi}}{\beta J \cosh \left( \frac{\pi t}{\beta}\right)}.
		\end{equation}
		Now, consider a subsystem of $2j$ Majorana fermions and define the spin states only on this subsystem. We assume that $2j<N$ and that $j$ is large  $1\ll \beta J\ll j$. We consider the case that $j$ of the spins are measured, meaning that the values of $s_k ~,~k=1,...,j$ are known.
		
		Next, construct a ``mass-term'' perturbation:
		\begin{equation}
		\label{MassTerm}
		V(t)=- i\epsilon J \sum_{k=1} ^{j} s_k \psi_{2k-1} \psi_{2k},
		\end{equation}
		where $\epsilon \ll 1$. Since (\ref{MassTerm}) is a sum of terms like (\ref{Ofdiagonal}), the average value of the perturbation in the state $|B_s (\beta)\rangle$ is:
		\begin{equation}
\label{average1}
		\langle V(t) \rangle =-i \epsilon J~ \frac{-i\sqrt{\pi}}{\beta J \cosh \left( \frac{\pi t}{\beta}\right)}~ j=- \epsilon j \frac{\sqrt{\pi}}{\beta  \cosh \left( \frac{\pi t}{\beta}\right)}.
		\end{equation}
		
		The fact that the average value of the perturbation does not vanish depends crucially on the knowledge of the state of the spins. In the state $\rho\propto I$ for which the spin values are unknown:
		\begin{equation}
		\langle \psi_{2k-1} \psi_{2k} \rangle =\text{tr}\left( \rho \psi_{2k-1} \psi_{2k} \right)\propto \sum_{\{s_i\}} s_i = 0.
		\end{equation}
		Therefore, $\langle V(t)\rangle =0$. Similarly, it holds for thermal states.
		
Equation~(\ref{average1}) leads to a correction to the Schwarzian action. Defining $\phi$ by the derivative of the reparameterization mode, $f'=e^{\phi}$, one obtains the total action (see Appendix A):
		\begin{equation}
		S[\phi]=\int dt \left( \frac{\dot{\phi}^2}{J}-J\left( e^{\phi}-\hat{\epsilon} e^{\phi/2}\right)\right).
		\end{equation}
		The EOM leads to an equation expressing energy conservation:
		\begin{equation}
		E=\frac{\dot{\phi}^2}{J}+J\left( e^{\phi}-\hat{\epsilon} e^{\phi/2}\right).
		\label{e-conserve}
		\end{equation}
				
The initial condition is $\dot{\phi}(t=0)=0$. The energy is a sum of a ``kinetic-term'' and a Liouville-type potential. The potential is depicted in Figure 2 for three cases of varying perturbation strength: (i) unperturbed $\hat{\epsilon}=0$, (ii) small perturbation $\hat{\epsilon} \beta J \ll 1$ and  (iii) large perturbation $\hat{\epsilon} \beta J > \pi$.
		
		\begin{figure}[t]
			\begin{center}
				\includegraphics[scale=0.35]{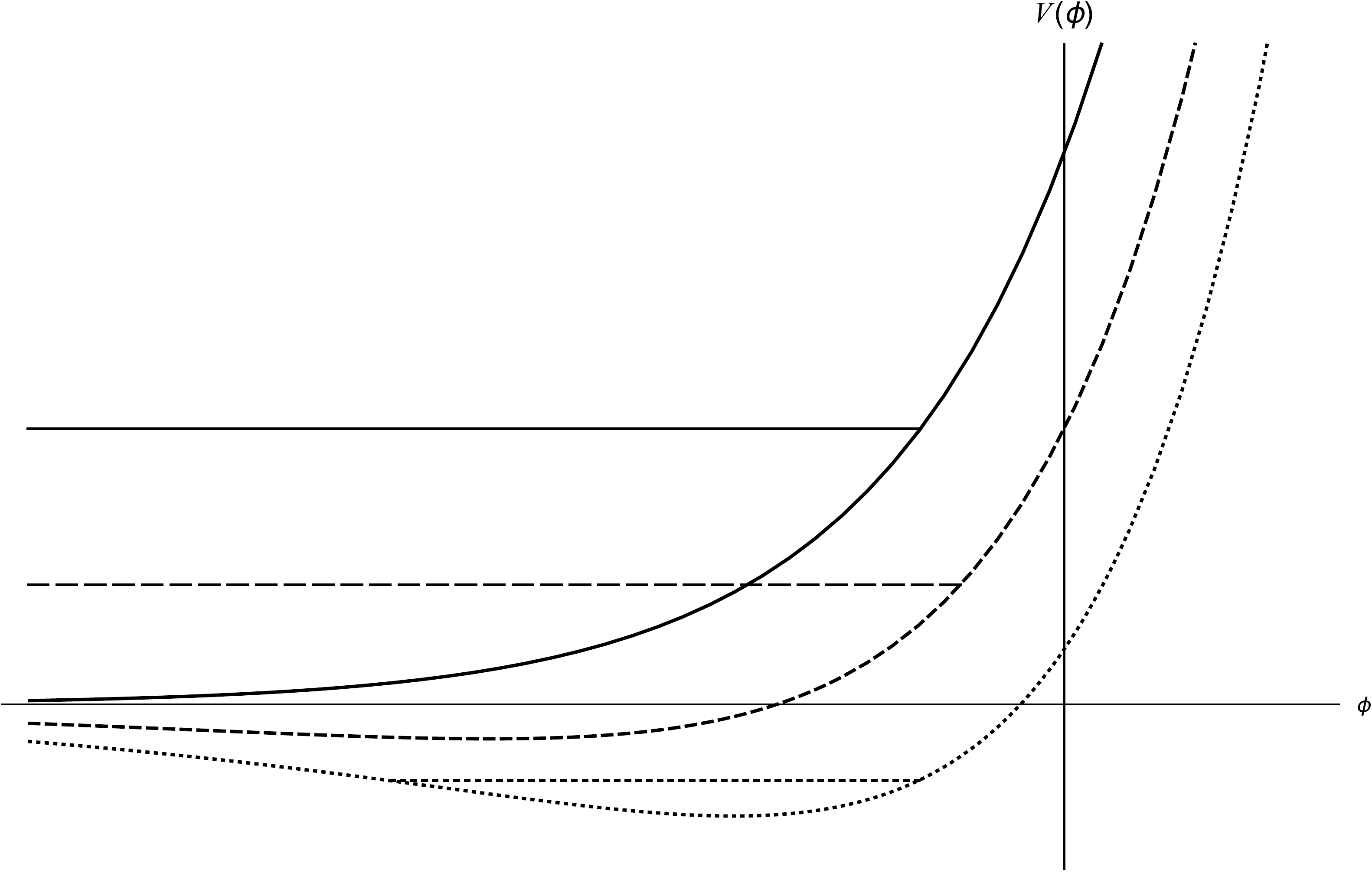}
			\end{center}
			\caption{Shown are: the unperturbed potential (solid), the potential perturbed by a small perturbation (dashed) and the potential perturbed by a ``large'' perturbation (dotted). The horizontal lines show the corresponding energies.  }
		\end{figure}
		
		Using $f'(0)= e^{\phi_0}$, we find that
		\begin{equation}
		\label{modenergy}
		E = J \left(e^{\phi_0} - \hat{\epsilon} e^{\phi_0/2} \right)=J \frac{\pi}{\beta J} \left( \frac{\pi}{\beta J}-\hat{\epsilon} \right).
		\end{equation}
		In Appendix B we calculate in detail the perturbed solutions for $z\sim f'=e^\phi$ and $T\sim f=\int dt e^{\phi}$, when $\hat{\epsilon}\beta J \ll 1$ .
		
		We can now state the main result of the paper: when the BH is perturbed, a region that was previously behind the horizon becomes exposed to an outside observer. The amount of exposure in global coordinates is proportional to the strength of the perturbation:
		\begin{equation}
		\frac{\Delta \tau}{\tau_0}=\frac{\pi-2}{4\pi} \hat{\epsilon} \beta J ,
		\end{equation}
		where $\tau_0 = \frac{\pi}{2}$.
		The numerical coefficient in front of the small parameter is positive: the observer will see more with respect to what she has seen in the absence of the perturbation.
		
		The new BH horizon is given by the line $\tau=\rho+\Delta \tau$, which is a null geodesic that the observer sees at the end of her trajectory, as depicted in Figure 3.
		
		\begin{figure}[t]
			\begin{center}
				\includegraphics[scale=0.35]{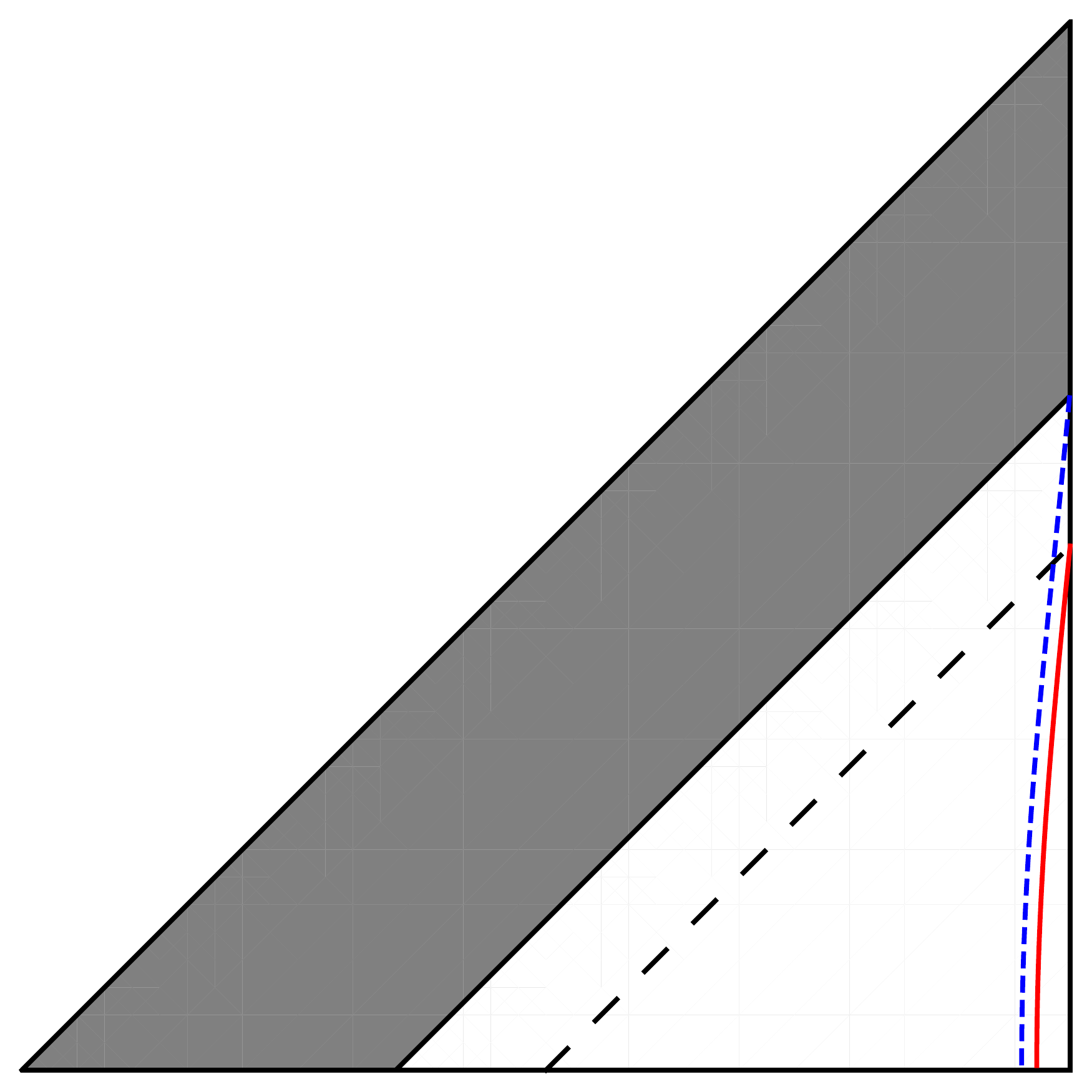}
			\end{center}
			\caption{The perturbation allows an observer to see a part of the interior of the unperturbed BH. A red line denotes the ``old'' trajectory while the blue-dashed line represents the ``new'' trajectory. The ``old'' horizon is the black dashed line while the ``new'' horizon is the black solid line, depicting the trajectory of a light ray leaving from the last point in the new boundary trajectory.}
		\end{figure}

		Next, we relate the degree of excitation to the degree of exposure. From Eq.~(\ref{modenergy}),
		\begin{equation}
		\frac{\Delta E}{E} =-\frac{\hat{\epsilon} \beta J}{\pi}
		\end{equation}
		It follows that:
		\begin{equation}
		\frac{\Delta \tau}{\tau_0} = \frac{\pi-2}{4} \frac{|\Delta E|}{E}.
		\end{equation}
		
We conclude that a measurement of $j$ spins corresponds to a small negative energy perturbation which leads to a small region of the BH interior being exposed. The extent of exposure is proportional to the number of measured spins, and so also to the strength of perturbation.

We now show that a measurement of $j$ spins is interpreted at the boundary not only as the exposure of a region behind the horizon but also as a burst of radiation coming from this region.  We show that if the BH is in equilibrium no radiation flows towards the boundary, while, in contrast, when the BH is out of equilibrium an energy flux does reach the boundary in the form of radiation.

The energy flux $F$ emitted into the bulk and arriving at the boundary is given by \cite{malda1} (We work with Lorenzian signature which results an additional minus sign with respect to \cite{malda1}),
$
F(t)=\frac{\bar{\phi}_r}{8\pi G} \frac{d}{dt} \{T(t),t \}.
$
In SYK terms,
\begin{equation}
\label{flux}
F(t)=\frac{N \alpha_S}{J}\frac{d}{dt} \{f(t),t \}.
\end{equation}
For a BH in equilibrium, $f(t)\propto \tanh(\omega t)$, which leads to $F(t)=0$ and therefore to a vanishing energy flux at the boundary.

Next, consider the measurement of $j$ Majorana spins and the ``mass-term'' perturbation. From Eq.~(\ref{Omega}) in Appendix B, $\omega=\frac{\pi}{2\beta}\sqrt{1-\frac{\hat{\epsilon} \beta J}{\pi}}$ and for the case $\hat{\epsilon}\beta J < \pi$, one obtains from Eq.~(\ref{zoft}),
\begin{equation}
\label{fprime}
f'(t)= \frac{(\pi-\hat{\epsilon}\beta J)^2}{\left(-\hat{\epsilon}\beta J+(2\pi-\hat{\epsilon}\beta J)\cosh(\omega t) \right)^2}.
\end{equation}
Substituting Eq.~(\ref{fprime}) into Eq.~(\ref{flux}) yields the following ``dimensionless flux'',
\begin{equation}
\beta^2 F(t)=\frac{\pi^{\frac{3}{2}}}{4}\frac{\alpha_S N}{\beta J}\hat{\epsilon}\beta J\frac{(\pi-\hat{\epsilon}\beta J)^{\frac{3}{2}} (2\pi-\hat{\epsilon}\beta J) \sinh(\omega t) }{\left( (2\pi-\hat{\epsilon}\beta J)\cosh(\omega t)-\hat{\epsilon}\beta J \right)^2}.
\end{equation}

Since $F(t)>0$, the energy flows to the boundary in the form of radiation. Moreover, $\beta^2 F(t)$ is amplified by the large number $N \alpha_S/(\beta J)$.

For large times and $\hat{\epsilon} \beta J \ll 1$
\begin{equation}
\beta^2 F(t) = \frac{\pi^2}{4}\, \frac{\alpha_S N}{\beta J}\, \hat{\epsilon}\beta J\, e^{-\omega t},
\end{equation}
so the flux decays in a characteristic time $\omega^{-1}=\frac{2\beta}{\pi \sqrt{1-\frac{\hat{\epsilon}\beta J}{\pi}}}$ which increases as the strength of the perturbation increases.

\begin{figure}
\begin{center}
\includegraphics[scale=0.95]{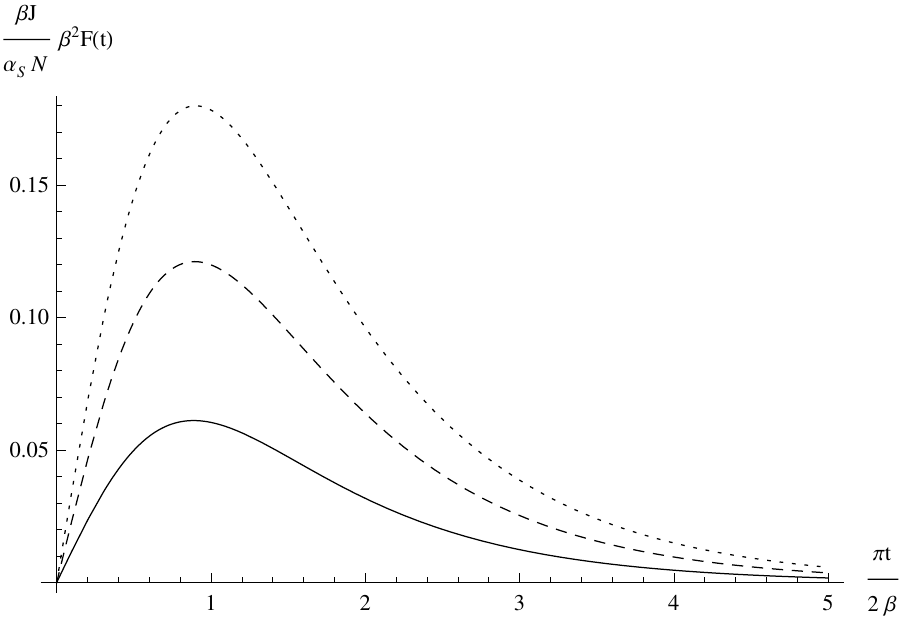}
\end{center}
\caption{The (dimensionless) energy flux into the boundary (times the factor $\beta J/N \alpha$) is depicted as a function of $\frac{\pi}{2\beta} t$, for three values of perturbation strengths: $\hat{\epsilon}\beta J =0.1,0.2,0.3$ (solid, dashed and dotted, respectively).}
\label{Flux123}
\end{figure}

In Figure \ref{Flux123}, $\beta ^2 F(t)$ is plotted as a function of the dimensionless boundary time $\pi t/(2 \beta)$ for three values of $\hat{\epsilon}\beta J$, $\hat{\epsilon}\beta J=0.1,0.2,0.3$. The boundary observer can interpret this radiation as arriving from the region that was behind the horizon of the BH. A phenomenon that does not occur according to classical GR.

		\section{Summary, interpretation and discussion}
		
		Using the SYK model to describe BHs in NAdS${}_2$ gravity, we showed explicitly how a negative-energy time-dependent perturbation allows an external observer to view regions of spacetime that were behind the horizon of the unperturbed BH.

Knowing the state of $j$ spins out of $N$, we were able to construct a ``mass-term'' perturbation resulting in a change in the potential of the reparametrization mode by a magnitude $\Delta E \propto j$. The perturbation shifts the horizon of the BH inwards, thus exposing to an external observer a region in spacetime that used to be inside the horizon.

From the boundary, this is viewed as a burst of radiation that seems to  arrive from the exposed region.  This result supports the general arguments presented in \cite{QuantumHair} that external perturbations which take BHs out of equilibrium can reveal their interior via the properties of the emitted radiation while they relax to equilibrium. This idea can be perhaps verified by the detection of specific gravitational wave emission from astrophysical BH collisions, in addition to the emission predicted by GR.

		\section*{Acknowledgments}
		We would like to thank Juan Maldacena for discussions about his work and for comments on the manuscript. The research of RB and YZ was supported by the Israel Science Foundation grant no. 1294/16.

	\end{document}